\documentclass[11pt]{article} % smaller base font

\usepackage[T1]{fontenc}
\usepackage[utf8]{inputenc}
\usepackage{amsmath, amssymb}
\usepackage{graphicx}
\usepackage{cite}
\usepackage{hyperref}
\usepackage[top=1in, bottom=1in, left=0.9in, right=0.9in]{geometry} % tighter margins
\usepackage{titlesec} % control heading size/spacing

\titleformat{\section}{\normalfont\normalsize\bfseries}{\thesection}{1em}{} 
\titleformat{\subsection}{\normalfont\normalsize\itshape}{\thesubsection}{1em}{} 
\titlespacing*{\section}{0pt}{1ex plus .5ex minus .2ex}{0.8ex}
\titlespacing*{\subsection}{0pt}{0.8ex plus .4ex minus .2ex}{0.6ex}

\title{Nanoscale mapping of stacking-dependent work function and local photoresponse in CVD-grown MoS$_2$ bilayers by KPFM}
\author{
  Anagha Gopinath$^{1}$,
  Faiha Mujeeb$^{2}$, 
  Subhabrata Dhar$^{2}$,
  Jyoti Mohanty$^{1}$\thanks{Email: jmohanty@phy.iith.ac.in}   
}

\date{}
\begin{document}
\maketitle
\begin{center}
$^{1}$Nanomagnetism and Microscopy Laboratory, Department of Physics, Indian Institute of Technology Hyderabad, Kandi, Sangareddy, Telangana, India \\
  $^{2}$Department of Physics, Indian Institute of Technology Bombay, Powai, Mumbai, India\\
\end{center}
\begin{abstract}
Stacking order in bilayers of transition metal dichalcogenides (TMDs) controls structural symmetry and layer-to-layer interactions, offering a direct route to tune their electronic properties and enable optoelectronic applications. The work function is a key parameter that determines the electronic and optoelectronic device performance. However, a comprehensive understanding of the influence of stacking order on work function of TMDs remains limited. Herein, we employ Kelvin Probe Force Microscopy (KPFM) to probe spatial variations in surface potential and thereby determine the work function of AA'- and AB-stacked MoS$_2$ bilayers grown using NaCl-assisted chemical vapor deposition (CVD) technique. The work function increases with layer number in both AA'- and AB-stacked MoS$_2$, with a larger work function difference in AB-stacked layers, reflecting their stronger interlayer coupling. KPFM measurements clearly resolve local electronic heterogeneities arising from carrier trapping at residual surface particulates from CVD growth. Photoinduced surface potential variations imply n-type doping in MoS$_2$ due to enhanced photogating from trapped holes and Na$^+$ ions at the MoS$_2$/SiO$_2$ interface. Our study demonstrates the competing effects of interlayer coupling, substrate-induced photogating, and carrier trapping by surface particulates in determining the localized optoelectronic response of MoS$_2$ bilayers. Correlative atomic force microscopy measurements in lateral force microscopy and force modulation microscopy modes probe the nanomechanical response to electronic variations. These findings provide new insights into the localized optoelectronic response of CVD-grown AA'- and AB-stacked MoS$_2$, with significant implications for the design and reliability of optoelectronic devices.
\end{abstract}

\section{Introduction}
Transition-metal dichalcogenides have emerged as a promising class of two-dimensional materials for nanoelectronic and optoelectronic applications~\cite{wang2012electronics,radisavljevic2011single,yin2012single,wu2014piezoelectricity}, owing to their unique, layer-dependent electronic~\cite{singh2022origin,kim2015engineering}, optical~\cite{lezama2015indirect,li2017layer} and mechanical properties~\cite{falin2021mechanical}. Monolayer TMDs exhibit a direct band gap~\cite{mak2010atomically,splendiani2010emerging}, strong spin-valley coupling~\cite{zhu2011giant,zeng2012valley,mak2012control}, robust light-matter interactions~\cite{mak2018light,liu2015strong}, and high carrier mobility~\cite{shi2021superior,majumder2025unveiling}. Beyond monolayers, layer-by-layer vertical integration of these TMDs enable the engineering of van der Waals homojunctions. Due to the higher density of states,  enhanced carrier mobility and better chemical stability compared to monolayers~\cite{late2013sensing,zhang2019transition}, few-layered TMDs, particularly bilayers, offer strong potential for applications in sensors~\cite{li2012fabrication}, photodetectors~\cite{xu2016atomically}, and field effect transistor~\cite{bao2013high}. The twist angle between adjacent layers introduces an additional degree of freedom in such stacked systems~\cite{fox2023stacking}. This unique structural tunability has been shown to drive the emergence of remarkable physical effects such as Moir\'{e} excitons~\cite{andersen2021excitons}, unconventional superconductivity~\cite{xia2025superconductivity,lin2018moire}, ferroelectricity~\cite{weston2022interfacial}, Mott insulator~\cite{wang2020correlated}, interlayer exciton emissions~\cite{barman2022twist}, non-linear optics~\cite{van2014tailoring,kim2024three}, etc., thereby opening up ways to explore strongly correlated electron systems and foster twistronics for advanced device applications.\\

The relative stacking configuration between adjacent layers governs the crystal symmetry and Coulombic interactions in bilayer TMDs~\cite{yan2015stacking,he2014stacking}, thereby influencing their optical and electronic properties~\cite{xia2015spectroscopic,he2014stacking,yan2016role}. Among them, the most energetically favorable polytypes are 2H (AB-stacking) and 3R (AA'-stacking) with trigonal prismatic coordination within the layer and distinct interlayer stacking order~\cite{xia2015spectroscopic}. These structural variations influence interlayer interactions, and stronger interlayer coupling in centrosymmetric 2H-MoS$_2$ has been shown to facilitate the formation of interlayer excitons~\cite{paradisanos2020controlling}. Zheng \textit{et al}. reported room-temperature ferroelectricity in 3R-MoS$_2$ driven by interlayer sliding~\cite{jiang2024vapor}. In contrast, vertical ferroelectricity was observed in 2H-MoS$_2$ through strain engineering~\cite{mao2024strain}. Recently, enhanced photoresponsivity has been reported in 2H-MoS$_2$ compared to 3R-MoS$_2$.~\cite{aggarwal2025stacking}. Thus far, chemical vapor deposition (CVD), using NaCl as a growth promoter, has been widely employed to obtain these AA'- and AB-stacked TMD domains~\cite{xu2024reconfiguring,hong2025strain,yan2025preparation}. Despite its potential to increase lateral domain size and enable large-scale growth of TMDs, NaCl-assisted growth leaves behind significant surface residual particles and adsorbates~\cite{singh2021nacl,zhang2017damage,jiang2022self}. Zhang \textit{et al}. reported a 20x reduction in photoluminescence intensity and a decline in transistor performance in MoS$_2$ grown using NaCl-assisted CVD growth, with a two-order-of-magnitude reduction in ON/OFF ratio and decreased field-effect carrier mobility~\cite{zhang2018considerations}. These findings underscore the need for a more detailed understanding of the effects of the surface residues on the electronic properties of TMDs, especially given the widespread use of the NaCl-assisted method to develop MoS$_2$ stacks for optoelectronic applications~\cite{luo2021layer,wang2022fast,fan2024layer} which demand materials of superior quality. Kelvin probe force microscopy (KPFM), a scanning probe microscopy technique, enables the spatially resolved evaluation of electronic properties, such as nanoscale mapping of surface potential and the corresponding work function~\cite{gu2023twist,lavini2018friction}. Previous KPFM studies in TMDs have revealed a correlation between layer thickness and work function~\cite{kaushik2015nanoscale,choi2014layer}. Furthermore, Kim \textit{et al}. probed charge transfer in monolayer TMDs induced by molecular adsorption, highlighting the potential of the technique in mapping the surface potential with high spatial resolution and sensitivity~\cite{kim2015work}. Yim \textit{et al.} directly visualized the light-induced surface potential shifts of monolayer MoS$_2$~\cite{yim2022imaging}. In a recent study, the optoelectronic response of MoS$_2$ bilayers with a twist angle of 0.17$^\circ$ was investigated using photosensitive frequency-modulated KPFM~\cite{zhao2025localized}, revealing periodic superstructures under dark conditions. Although the localized electronic and optoelectronic response of monolayers and twisted layers of TMDs has been explored using KPFM, the role of intrinsic stacking order on the local surface potential and its evolution under light illumination remains elusive till date. This is particularly important as small twist angles allow lattice reconstruction into stable 2H and 3R stacked configurations~\cite{weston2020atomic}. Furthermore, the impact of surface residues introduced during NaCl-assisted CVD growth of TMDs on local electronic and optoelectronic responses has not yet been explored. Understanding the role of stacking order and the impact of growth-induced residues on the local optoelectronic behaviour of TMDs is pivotal for optimizing these materials for future advanced device applications.\\

In the present work, we investigate the influence of stacking order and surface particles on the local surface potential and work function of NaCl-assisted CVD-grown MoS$_2$ bilayers using KPFM, focusing on the energetically stable AA' and AB stacking configurations. External illumination at a wavelength of 633 nm was employed to examine the photo-induced tunability of the surface potential. Raman spectroscopy, photoluminescence (PL) spectroscopy, and X-ray photoelectron spectroscopy (XPS) provide insight into the corresponding structural, optical, and chemical properties. Complementary lateral force microscopy (LFM) and force modulation microscopy (FMM)  measurements further reveal the nanoscale mechanical characteristics across various morphologies. Our findings elucidate the local optoelectronic response of AA'- and AB-stacked MoS$_2$ and establish the roles of stacking configuration and local heterogeneitieson the electronic properties of MoS$_2$.

\section{Experimental Techniques}
\subsection{CVD growth of AA'- and AB-stacked MoS$_2$}
 MoS$_2$ was grown on SiO$_2$/Si substrate using the chemical vapor deposition (CVD) technique in a horizontal tube furnace with two heating zones~\cite{anagha2025unveiling}. Fig.~\ref{Figure1}a represents the schematic of the experimental setup. The substrates were sonicated sequentially in acetone, isopropyl alcohol (IPA) and deionized water for 4 minutes each and then blow-dried using Nitrogen. An alumina boat containing 70 mg of NaCl and 2.5 mg of MoO$_3$ was positioned in the second heating zone. Another boat containing 150 mg of Sulfur was placed in the first heating zone. The cleaned substrates were kept, with one face-up next to the MoO$_3$ and NaCl mixture in the alumina boat to promote the growth of AA'- and AB-stacked MoS$_2$, and the other face-down on top of the alumina boat to obtain monolayer MoS$_2$. To form an inert atmosphere, the system was flushed using high-purity Argon gas at a flow rate of 300 sccm. The temperature of the Mo and S precursor zones was ramped to 750$^\circ$C and 170$^\circ$C, respectively, in 50 minutes and maintained at these temperatures for 6 minutes. Argon gas, acting as the carrier gas, was supplied at a flow rate of 120 sccm until a temperature of 700$^\circ$C was reached in the Mo precursor zone, and then the flow was reduced to 70 sccm. Following the growth process, the system was allowed to naturally cool down to room temperature.

\subsection{Kelvin probe force microscopy (KPFM)}
KPFM measurements were performed using the Park Systems NX10 model. A Cr/Au-coated Silicon tip with a force constant of 5 N/m and a resonance frequency of 160 kHz was used. An AC voltage was applied to the tip at a frequency of 3 kHz with an amplitude of 4 V. A feedback system was employed to nullify the potential difference between the tip and the sample by applying a DC bias, which is equivalent to the surface potential of the sample. The contact potential difference is related to the work function as follows~\cite{gopinath2026optically}:
\begin{equation}
e V_{CPD} = \phi_t - \phi_s
\end{equation}
where $\phi_t$ and $\phi_s$ are the workfunction of the tip and sample respectively and V$_{CPD}$ is the contact potential difference between tip and sample~\cite{gopinath2026optically}. Calibration of the tip's work function was performed using the HOPG sample as a reference. To probe the light-dependent behavior, an external laser diode (Thor Labs) with a wavelength of 633 nm and a power of 5 mW was utilized to illuminate the sample during the measurements. 

\subsection{Lateral force microscopy (LFM) and Force modulation microscopy (FMM)}

LFM and FMM measurements were performed in contact mode using a cantilever with a backside reflective coating of Al, a spring constant of 2.8 N/m, and a resonance frequency of 75 kHz in the Park Systems NX10 Model. Lateral force maps were obtained by assessing the lateral deflection as the tip was moved forward and backward. For FMM measurements, additional mechanical oscillations were induced in the cantilever by applying a sinusoidal drive signal of 152 kHz to the z-piezo of the AFM. Surface properties were extracted by analyzing the acquired FMM phase and amplitude maps.
\subsection{Raman spectroscopy, Photoluminescence spectroscopy, X-ray photoelectron spectroscopy}

Raman Spectroscopy and Photoluminescence spectroscopy were conducted using a micro-PL/Raman system (Renishaw Invia Reflex, UK). The measurements were performed with 532 nm laser excitation and 0.5 mW laser power, using a 50x objective. A grating of 2400 lines/mm and 600 lines/mm was utilized for Raman and PL measurements, respectively. XPS measurements were done using a monochromatic Al-K$\alpha$ X-ray source with an energy of 1.486keV in an AXIS Supra system (Kratos Analytical).

\section{Results and Discussions} 
\begin{figure*}[h!]
  \centering
  \includegraphics[width=0.98\linewidth]{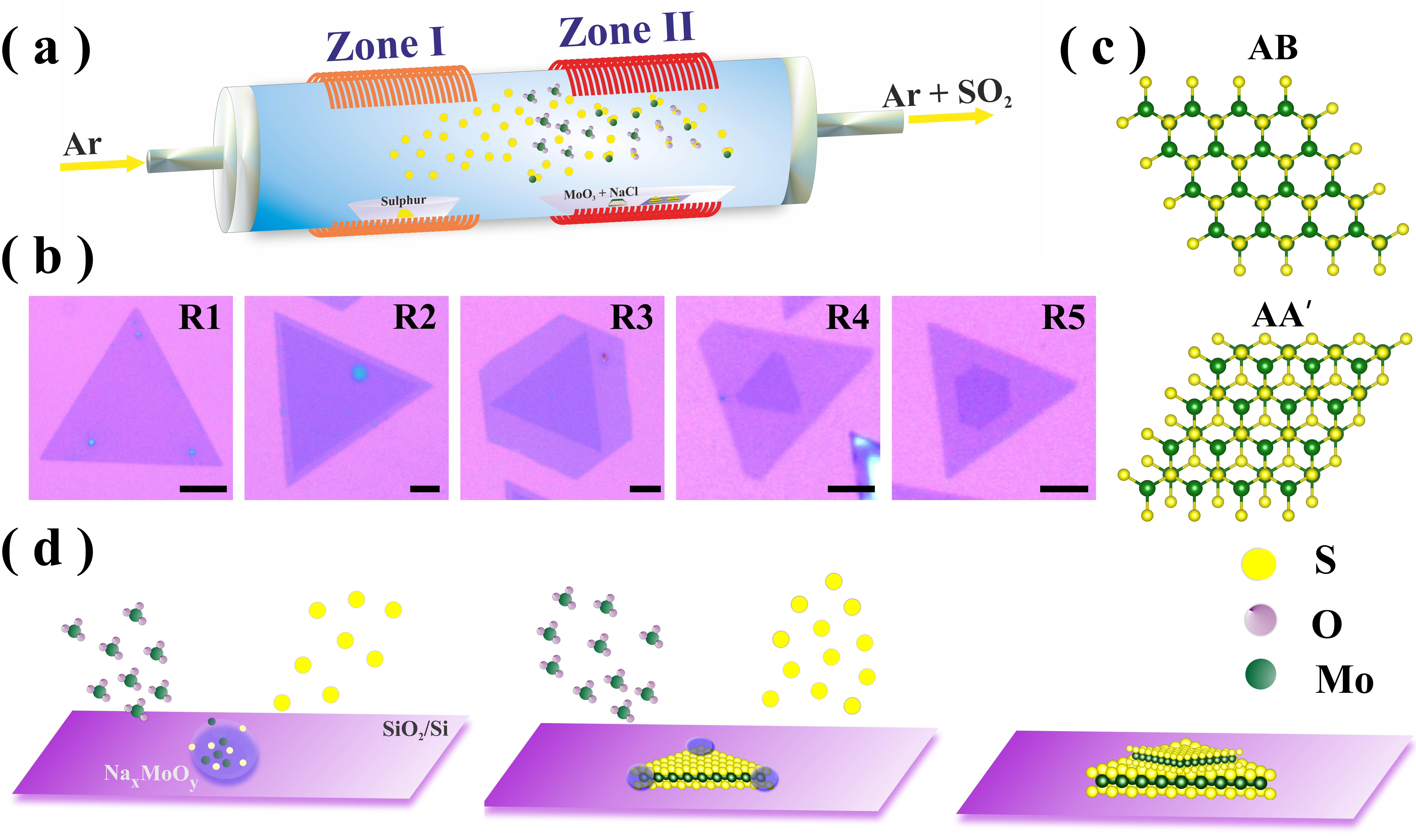}
  \caption{CVD growth of MoS$_2$ nanostructures. (a) Schematic illustration of the CVD set-up used for the growth process. (b) Optical microscopy images of as-grown MoS$_2$ showing different morphologies, labeled as R1, R2, R3, R4, and R5, obtained under the specified growth conditions. The scale bar corresponds to 5$\mu$m. (c) Top view representations of MoS$_2$ layers highlighting AB stacking (2H) and AA' stacking (3R) configurations. (d) Schematic depiction of NaCl-assisted CVD growth mechanism of MoS$_2$ on SiO$_2$/Si substrate.}
  \label{Figure1}
\end{figure*}

The NaCl-assisted chemical vapor deposition technique was employed to fabricate MoS$_2$ nanostructures on SiO$_2$/Si as described in the experimental section. Optical microscopy images of the as-grown sample shown in Fig.~\ref{Figure1}(b) reveal uniform triangular-shaped MoS$_2$ flakes (region R1) and layered MoS$_2$ domains exhibiting two distinct relative crystallographic orientations with rotationally aligned layers (regions R2, R3, and R5) and layers with an interlayer twist angle of 60$^\circ$ (region R4), corresponding to the 3R (AA'-stacking) and 2H (AB-stacking)~\cite{xia2015spectroscopic,shinde2018stacking,yan2025preparation,mccreary2022stacking,fan2024layer,dong2024strain,luo2021layer}, respectively (Fig.~\ref{Figure1}(c)). Morphology of the as-grown MoS$_2$ nanostructures is significantly influenced by the presence of NaCl along with MoO$_3$ during the growth process, which in turn enhances the reaction rate. The growth occurs via vapor-liquid-solid (VLS) growth mechanism~\cite{li2018vapour} in which liquid-like Na$_x$MoO$_y$ droplets form on the SiO$_2$/Si substrate through the reaction of MoO$_3$ and NaCl. Gaseous Sulfur reacts with these intermediates and initiates MoS$_2$ nucleation and their lateral growth through the in-plane propagation of droplets~\cite{hong2025strain,yang2025operando,tyagi2025systematic}. A relatively high NaCl content compared to MoO$_3$ (NaCl:MoO$_3$-28:1) results in higher nucleation density over the substrate, causing faster nucleation and slower diffusion. Hence, the lateral growth is suppressed, and the subsequent nucleation occurs on the already existing MoS$_2$ flakes~\cite{aggarwal2022centimeter}. This drives the formation of vertical MoS$_2$ nanostructures. Schematic of the growth process is illustrated in Fig.~\ref{Figure1}(d).\\
\begin{figure*}[h!]
  \centering
  \includegraphics[width=0.98\linewidth]{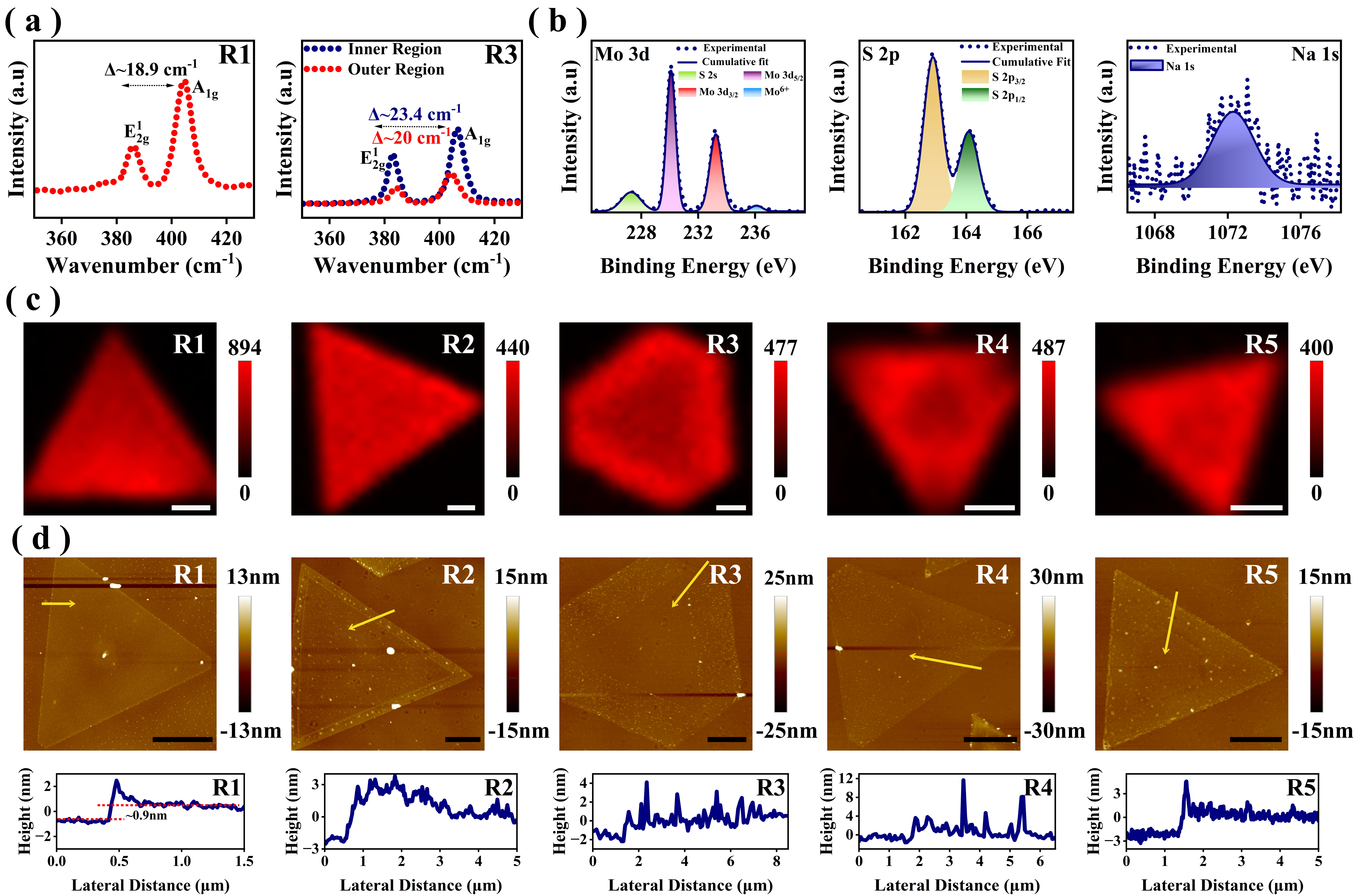}
  \caption{Characterization of as-grown MoS$_2$. (a) Raman spectra of region R1 and R3. $E_{2g}^{1}$ mode corresponds to in-plane vibration of Mo and S atoms, and the A$_{1g}$ mode corresponds to out-of-plane vibration of S atoms. For region R3, the red curve corresponds to data obtained from the outer region, and the blue curve represents the data from the inner region. (b) Core level XPS spectra of Mo 3d, S 2p, and Na 1s. (c) PL mapping of A-exciton intensity of regions R1-R5. Scale bar corresponds to 5$\mu$m. (d) Atomic force microscopy (AFM) images showing surface topography of regions R1-R5 and the corresponding line profiles. The yellow arrow in the topography map indicates the line along which the profiles are extracted. Scale bar corresponds to 4$\mu$m. }
  \label{Figure2}
\end{figure*}

Room-temperature Raman spectroscopy and PL spectroscopy measurements were employed to understand the structural and optical properties of as-grown MoS$_2$ nanostructures. Fig.~\ref{Figure2}(a) shows the Raman spectra of regions R1 and R3 and the spectra of regions R2, R4, and R5 are provided in the supporting information as Fig. S1. The Raman spectra exhibit two characteristic modes of MoS$_2$, the in-plane $E_{2g}^{1}$ mode and the out-of-plane A$_{1g}$ mode. A peak separation of $\sim$18.9 cm$^{-1}$ between $E_{2g}^{1}$ mode ($\sim$385.6 cm$^{-1}$) and $A_{1g}$ mode ($\sim$404.5 cm$^{-1}$) is observed for region R1, consistent with the Raman signature of monolayer MoS$_2$~\cite{majumder2025unveiling}. However, the inner regions of flakes R2-R5 exhibit a redshift of the $E_{2g}^{1}$ mode and a blueshift of the A$_{1g}$ mode relative to the corresponding modes in the outer region. This results in a larger peak separation in the inner region ($\sim$23 cm$^{-1}$) than the outer ($\sim$20 cm$^{-1}$), indicating the bilayer nature of MoS$_2$ in line with the previous reports~\cite{fan2024layer,hamza2022cvd}. The observed shifts arise from the enhanced interlayer vdW coupling that increases the restoring force and stiffens the A$_{1g}$ mode, whereas stronger dielectric screening weakens long-range Coulombic interactions and redshifts the $E_{2g}^{1}$ mode~\cite{aggarwal2022centimeter,li2012bulk}. The PL spectrum of region R1 (Fig. S2(a) in the supporting information) shows strong A excitonic emission at $\sim$1.82 eV accompanied by a weak B excitonic peak at $\sim$1.98 eV denoting the direct band-gap transitions at the K-point, further supporting the monolayer behavior of region R1~\cite{mak2010atomically,splendiani2010emerging}. Fig.~\ref{Figure2}(c) presents the PL map of the A exciton intensity in which regions R2-R5 feature lower PL intensity than region R1. Moreover, stronger A exciton emission is evident in the outer regions of R2-R5 than in the corresponding inner regions. This aligns with the A exciton quenching seen in the PL spectra of inner regions of R2-R5 (Fig. S2(a) of supporting information), and suggests the suppression of direct excitonic recombinations, indicating layered MoS$_2$. PL intensity maps of the B exciton of regions R1 to R5 are included in the supporting information (Fig. S2(b)). To further examine the surface topography, atomic force microscopy (AFM) measurements were performed. The AFM topography maps and corresponding line profiles are shown in Fig.~\ref{Figure2}(d). The line profile of region R1 exhibits a thickness of $\sim$0.9 nm, characteristic of monolayer MoS$_2$~\cite{li2017layer}. Although flakes R2-R5 showcase distinct inner and outer regions, the AFM line profiles do not reveal a pronounced height step at their boundary. The apparent height measured from the substrate to the centre of the flake is larger than the reported thickness of bilayer MoS$_2$, likely due to the residual growth intermediates formed during NaCl-assisted CVD growth  that appear as surface particles in the AFM image~\cite{suleman2022nacl,jiang2022self}. XPS characterization was employed to analyze the chemical states of the as-grown sample. Fig.~\ref{Figure2}{b} depicts the deconvoluted XPS spectra of Molybdenum (Mo) 3d and Sulphur (S) 2p core levels. In the Mo 3d core level spectrum, the dominant peaks at $\sim$230.09 eV and $\sim$233.24 eV originate from the Mo 3d$_{5/2}$ and Mo 3d$_{3/2}$ states of Mo$^{4+}$. The peak at $\sim$ 227.2 eV is associated with S 2s state and the faint shoulder at $\sim$236.1 eV indicates the Mo$^{6+}$ state from MoO$_3$ formation~\cite{kim2014influence}. The S$^{2-}$ peaks are detected at $\sim$162.9 eV (S2p$_{3/2}$) and $\sim$164.08 eV (S2p$_{1/2}$). In addition to the characteristic peaks of MoS$_2$, a distinct Na 1s peak is observed at $\sim$1072 eV~\cite{chen2021study}. This suggests that residual Na from the growth process remains on the surface of MoS$_2$, possibly in the form of growth intermediates~\cite{suleman2022nacl}.\\
\begin{figure*}[h!]
  \centering
  \includegraphics[width=0.98\linewidth]{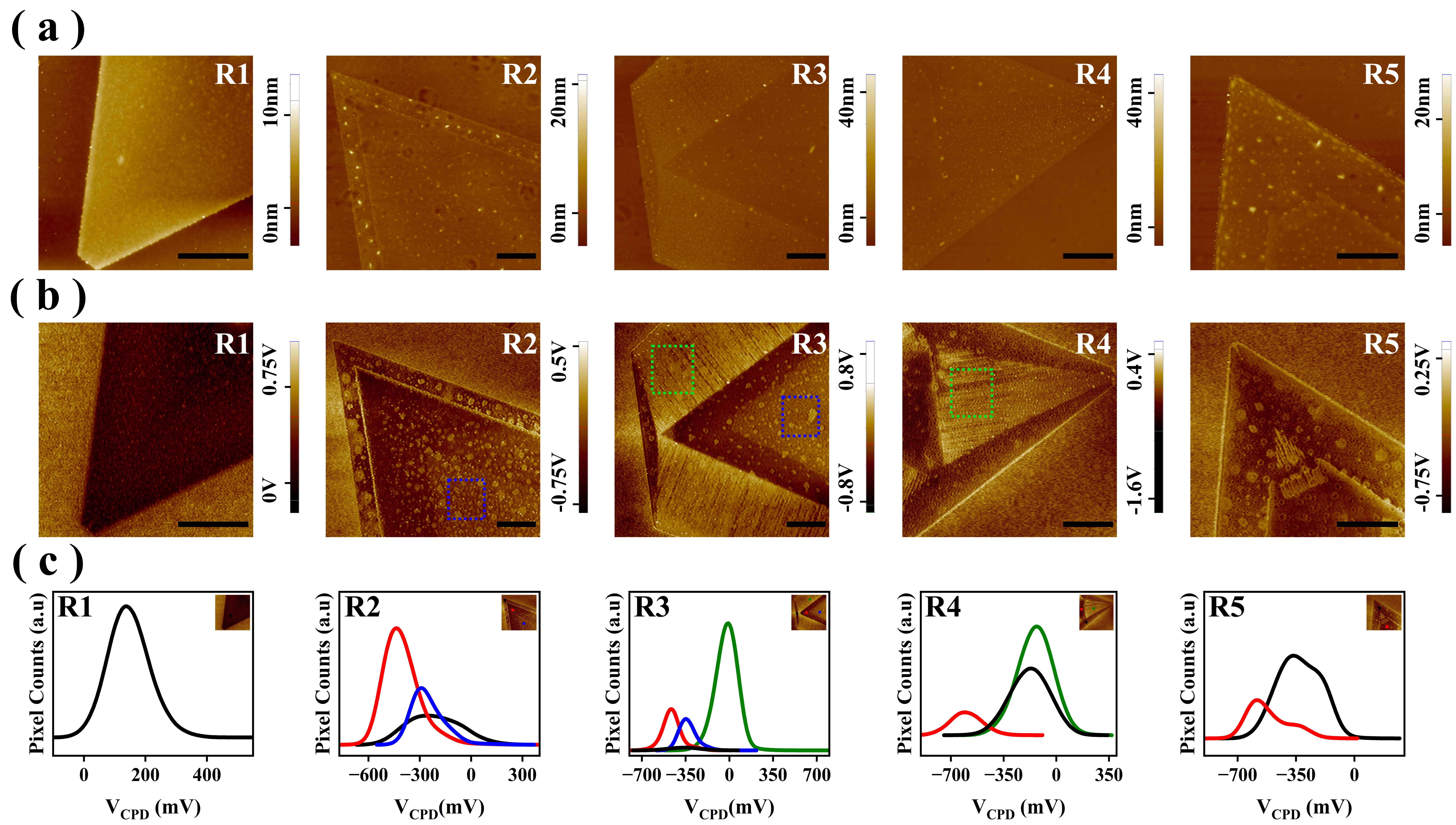}
  \caption{KPFM characterization of as-grown MoS$_2$. (a) Zoomed-in topography images of regions R1-R5. (b) Surface potential maps of regions R1-R5 obtained under dark conditions. Stripe-like features in regions R3 and R4 are highlighted with a green box, and distinct secondary contrast on the top layer of regions R2 and R3 are highlighted in blue. Scale bar corresponds to 2$\mu$m. (c) Surface potential histograms obtained from  KPFM maps. The regions from which the histograms are extracted are highlighted in the inset using corresponding colors.  }
  \label{Figure3}
\end{figure*}

\begin{figure*}[h!]
  \centering
  \includegraphics[width=0.98\linewidth]{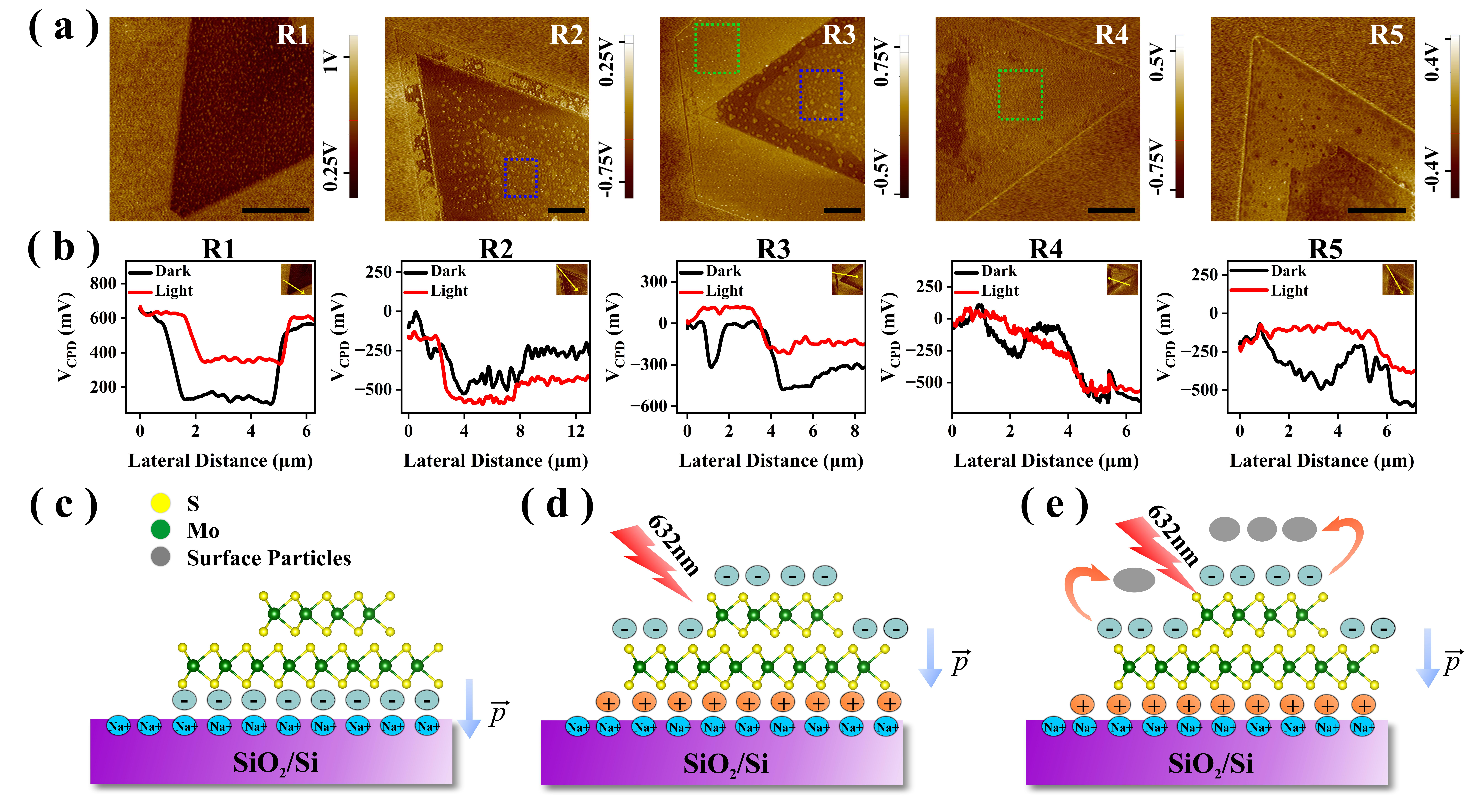}
  \caption{KPFM characterization of as-grown MoS$_2$ under 633 nm laser illumination. (a) Surface potential maps of regions R1-R5 obtained under laser illumination. Scale bar corresponds to 2$\mu$m. Stripe-like features in regions R3 and R4 are highlighted with a green box, and distinct secondary contrast on the top layer of regions R2 and R3 are highlighted in blue. (b) Surface potential profiles along the lines marked in the inset under dark and illuminated conditions. Schematic illustrating (c) interfacial dipole formation in MoS$_2$/SiO$_2$/Si, (d) photocarrier generation and photogating in MoS$_2$ under illumination, and (e) carrier trapping by the surface adsorbates.}
  \label{Figure4}
\end{figure*}

To elucidate the electronic properties of CVD-grown AA'- and AB-stacked MoS$_2$, Kelvin probe force microscopy (KPFM) measurements were performed. Prior to the measurements, the work function of the tip was calibrated using a standard HOPG sample. The measured contact potential difference (V$_{CPD}$) reflects the surface potential, arising from the workfunction difference between the tip and the sample. Fig.~\ref{Figure3}(a) shows the zoomed-in topography mappings of regions R1-R5. The spherical particulates are clearly evident at the edges of the inner and outer crystal domains of regions R2-R5 and also appear as an extended boundary layer along the edges of the inner crystal domains in regions R3 and R4. Such edge particles were reported previously in NaCl-assisted monolayer WS$_2$ growth and were identified as NaCl solid solutions dissolved by Na complexes formed with transition metal and chalcogen atoms~\cite{jiang2022self}. Consistent with this observation, XPS measurements revealed the presence of Na (Fig.~\ref{Figure2}(b)), suggesting that these particulates are associated with Na-derived species. Fig.~\ref{Figure3}(b) and Fig.~\ref{Figure3}(c) represent the surface potential mappings and respective histograms obtained under dark conditions using KPFM. The corresponding workfunction mappings are given in Fig. S3 of the supporting information. Monolayer region, R1, exhibits a relatively uniform surface potential with a workfunction of $\sim$5.0 eV. In contrast, pronounced spatial variations are apparent in surface potential mappings of regions R2-R5 distinct from R1, indicating local electronic heterogeneity. As evident from the surface potential histogram in Fig.~\ref{Figure3}(c), a decrease in surface potential is observed from the bottom to the top layer, as indicated by the black (bottom layer) and red (top layer) curves in the histogram, signifying a corresponding increase in the local work function (see Fig. S3). The observed trend in workfunction with respect to layer number is consistent with the previous reports~\cite{kaushik2015nanoscale,choi2014layer} and can be attributed to the interfacial dipole formed at the MoS$_2$/SiO$_2$  interface (Fig.~\ref{Figure4}(c)) due to the adsorption of Na$^+$ ions on the surface of SiO$_2$ during NaCl-assisted CVD growth~\cite{han2021interface,kaushik2020charge}. The dipole experiences weak screening from the bottom layer and strong screening towards the top, thereby resulting in a higher workfunction in the top layer and reduced surface potential variations. However, the magnitude of workfunction difference is significantly higher in the AB-stacked layers ($\sim$420 meV for R4) than the AA'-stacked region ($\sim$ 180 meV for R2, $\sim$ 120 meV for R3, and $\sim$ 210 meV for R5). This larger workfunction shift in AB-stacked layers relative to AA'-stacking is driven by the stronger interlayer coupling that enhances charge redistribution in AB-stacked MoS$_2$ layers~\cite{van2019stacking,fan2016modulation,paradisanos2020controlling} whereas weaker interlayer hybridization in AA'-stacking results in the smaller surface potential difference. Furthermore, surface particulates observed at the edges of the inner triangular domains in R3 and R4, appearing as the boundary layer, give rise to a distinct striped pattern in the surface potential mapping, as highlighted by the green box in regions R3 and R4 in Fig.~\ref{Figure3}(b). This region is characterized by a higher surface potential than the corresponding monolayer and bilayer regimes of R3 and R4 (green curve in the surface potential histogram in Fig.~\ref{Figure3}(c)). In addition, the spherical nanoscale particulates seen on the MoS$_2$ surface also exhibit distinct surface potential contrast in the KPFM map. Such changes can arise from charge transfer between these particles and MoS$_2$, interfacial dipole formation or from localized carrier trapping which can modify the local vacuum level and Fermi level alignment. The dense distribution of particulates extending from the inner triangular edge across the outer MoS$_2$ region in R3 and R4 results in overlapping electrostatic perturbations, giving rise to the stripe-like pattern in the KPFM maps.  In addition to the localized variations, within the inner triangular domains of R2 and R3, a distinct secondary contrast, as indicated by the blue box in regions R2 and R3 in Fig.~\ref{Figure3}(b), appears in the form of a triangular feature at a higher potential than the underlying bilayer region. The surface potential histogram associated with the observed contrast of regions R2 and R3 is represented by the blue curve in Fig.~\ref{Figure3}(c). This feature is not discernible in the corresponding topography maps and is likely due to the higher density of surface particulates at the centre of the inner triangular domain. Owing to the reduced screening in atomically thin MoS$_2$, the electrostatic perturbations are not confined to the immediate vicinity of these spherical particles and can extend laterally into the surrounding layer. Hence, the distinct triangular features of regions R2 and R3 point to the electrostatic response of the underlying layer to spatial variations in the particulate-induced charge distribution. The presence of these particulates contributes to the increased spread of the surface potential histograms and the appearance of multiple peaks in the surface potential histogram of region R5 and the outer region of R2 (indicated by the black curve).\\

In order to probe the localized optoelectronic response, KPFM measurements were performed with a 633 nm laser. Fig.~\ref{Figure4}(a) and \ref{Figure4}(b) represent the surface potential mappings of regions R1-R5 obtained under laser illumination and the corresponding KPFM line profiles derived from dark and light measurements, respectively. Surface potential histograms and the corresponding work function mappings are included in the supporting information as Fig. S4 and Fig. S5, respectively. The extracted work function values for regions R1-R5 under dark and illuminated conditions are presented in Table.1 of the supporting information. Based on the line profiles shown in Fig.~\ref{Figure4}(b), region R1 exhibits a $\sim$ 220mV increase in surface potential with a corresponding reduction in work function under illumination relative to the dark measurements. A similar photoinduced response is observed in the bottom layer of regions R2 to R5. These trends are consistent with the photogating mechanism. Upon illumination, electron-hole pairs are generated in MoS$_2$ since the laser photon energy ($\sim$ 1.96 eV) exceeds the band gap. The photoexcited holes are preferentially trapped at the MoS$_2$/SiO$_2$ interface, acting as a positive gate~\cite{yim2022imaging}. Together with Na$^+$ ions, they enhance the photogating effect by increasing the charge carrier density at the interface. Thus, the electrons remain on the MoS$_2$ surface, leading to an upward shift in the Fermi level and, consequently, n-type doping with a reduction in the work function. With the exception of region R2, the bilayer regions of flakes R3 to R5 demonstrate a similar photoresponse, consistent with the underlying monolayer (Fig.~\ref{Figure4}(b) and Fig. S5 of supporting information). However, the decrease in work function is less pronounced in the top layers of regions R3-R5 (see Table.1 in the supporting information), resulting in an increased work function contrast between the top and bottom layer under illumination relative to dark conditions. Stronger substrate-induced photogating in the bottom layer and reduced electrostatic modulation of the top layer due to enhanced screening and interlayer redistribution of photogenerated carriers govern the observed trend.  In contrast, the bilayer region of R2 shows an increase in work function from 5.26 eV to 5.39 eV upon illumination. This anomalous increase is likely due to the trapping of photogenerated electrons in localized defect states associated with growth-related residues or the surface adsorbates, which further lowers the Fermi level. However, the work function difference between the top and bottom layers remains higher in the AB-stacked configuration even under illumination, owing to enhanced charge redistribution between the layers resulting from stronger interlayer coupling. Accordingly, the observed variations in the surface potential of MoS$_2$ bilayers arise from the interplay between interlayer interactions, photogating induced by the substrate, and charge depletion governed by surface traps.  Moreover, illumination-induced charging effects are also evident in the particulate regions of R3 and R4. The trapping of photoexcited carriers by the surface particulates from the underlying MoS$_2$ layer drives local changes observed in the surface potential. It is noteworthy that the surface potential contrast between the stripes and the underlying MoS$_2$ in regions R3 and R4 fades under illumination (highlighted by the green box in Fig.~\ref{Figure4}(a)) as photogenerated carriers laterally redistribute in MoS$_2$  to screen the spatially varying electrostatic potential induced by the residue network. The secondary triangular contrast of regions R2 and R3 follows the photoresponse trend of the surrounding bilayer, further reaffirming its origin as the electrostatic response of the underlying layer to the localized surface potential variations of particulates. Schematics of the evolution of charge distribution under  illumination are shown in Fig.~\ref{Figure4}(c-e).\\  
\begin{figure}[h!]
  \centering
  \includegraphics[width=0.98\linewidth]{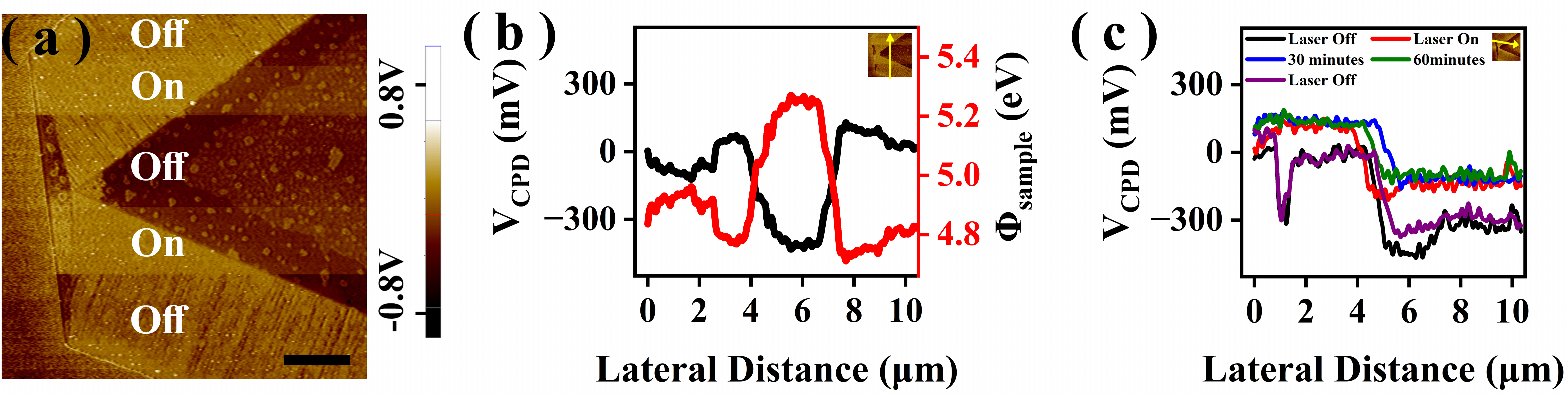}
  \caption{(a) Surface potential image of region R3 acquired by turning the laser ON and OFF in a single scan. Scale bar corresponds to 2$\mu$m. (b) Surface potential and work function profiles obtained along the line shown in the inset. (c) Surface potential profiles acquired at different times under laser illumination and after the laser was turned off, along the line shown in the inset.}
  \label{Figure5}
\end{figure}

To confirm the electronic origin of the observed changes, the laser illumination was turned on and off multiple times during a single scan of region R3. Switching between dark and illuminated states induces surface potential variations that appear as dark and bright bands in Fig~\ref{Figure5}(a).  The corresponding line profiles of surface potential and work function, with the work function map provided in Fig. S6(a), are depicted in Fig.~\ref{Figure5}(b). To evaluate the reproducibility and temporal response of surface potential, measurements were performed after 30 minutes and 1 hour of continuous laser illumination, and again after the laser was turned off. Line profiles acquired during the surface potential measurements are shown in Fig.~\ref{Figure5}(c) and the associated surface potential maps are provided in the supporting information (Fig. S6(b)). An enhanced surface potential is observed upon laser exposure as described in the previous section. However, no significant change is observed in the surface potential profiles recorded  after 30 minutes and 1 hour of laser illumination, indicating that the system has reached a stabilized photoinduced electrostatic state. Notably, the surface potential recovers to its initial state upon turning off the laser, consistent with photocarrier recombination and detrapping from the residues. The contrast between stripes and MoS$_2$ also reappears once the laser is turned off as the photoinduced charge redistribution relaxes and restores the initial electrostatic landscape (Fig.~\ref{Figure5}(a)). Reversibility of these surface potential variations confirms that the observed changes are electronic in origin and not a result of structural or chemical modifications, which is essential for stable and reliable optoelectronic device operation. However, non-uniform distribution of the surface particulates results in an inhomogeneous optoelectronic landscape. These features can locally modulate the photoresponse while compromising the uniformity and reliability of device performance.\\
\begin{figure*}[h!]
  \centering
  \includegraphics[width=0.98\linewidth]{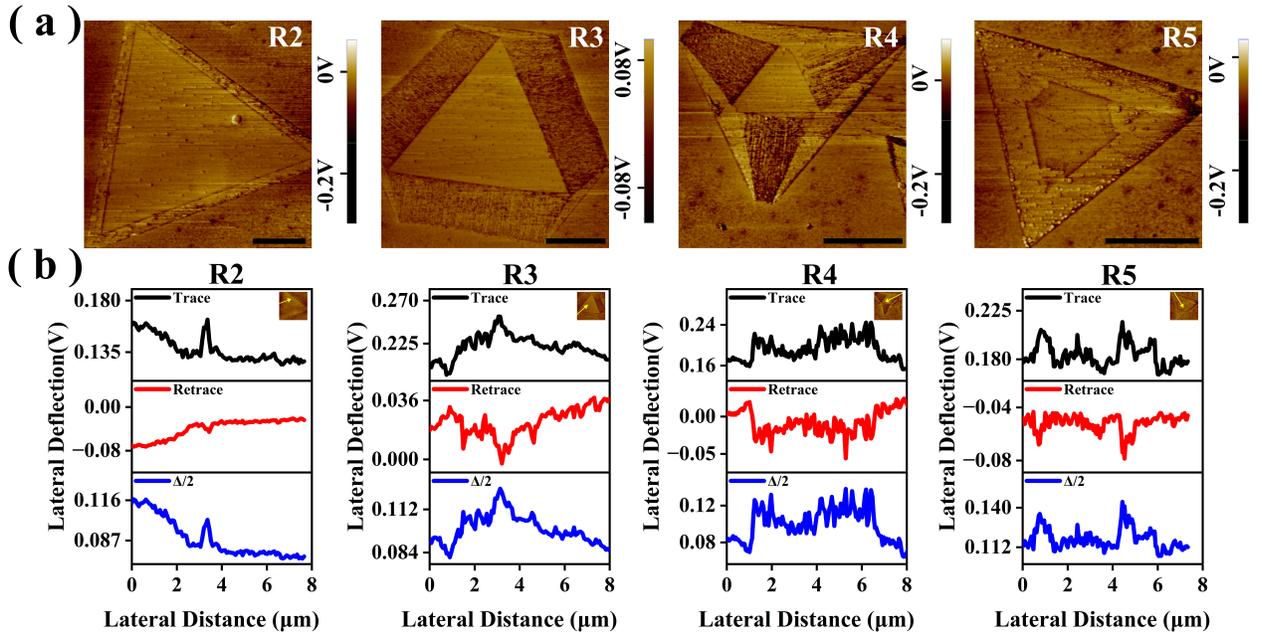}
  \caption{LFM characterization of regions R2-R5. (a) Torsion maps of regions R2-R5 in LFM mode. (b) Torsion profiles in trace and retrace mode, and their corresponding difference signal along the line shown in the inset. The difference signal confirms their frictional origin. Scale bar corresponds to 5$\mu$m.  }
  \label{Figure6}
\end{figure*}

\begin{figure}[h!]
  \centering
  \includegraphics[width=0.85\linewidth]{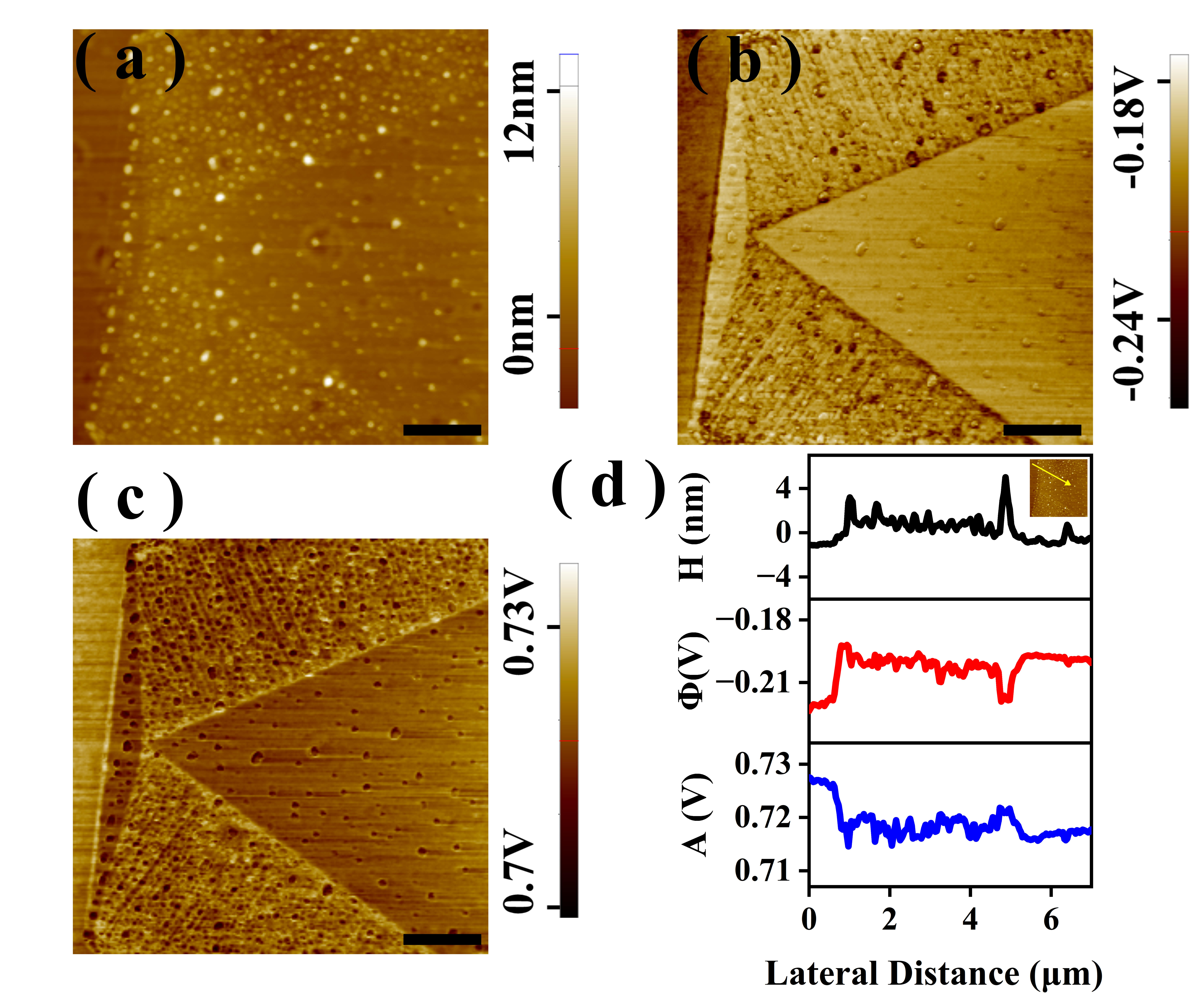}
  \caption{FMM characterization of region R3. (a) Topography, (b) Phase, and (c) Amplitude maps obtained in FMM mode and  (d) corresponding line profiles along the line shown in the inset. Here, H, $\Phi$, and A denote height, phase, and amplitude, respectively.  Scale bar corresponds to 2$\mu$m.}
  \label{Figure7}
\end{figure}

Since stripe domains in regions R3 and R4 and the secondary contrast within the bilayer of regions R2 and R3 revealed by KPFM mapping are not apparent in optical microscopy or PL mapping and show no Raman signatures, LFM and FMM measurements were performed to investigate their possible structural, mechanical, or electrostatic origins. LFM maps of regions R2-R5 are displayed in Fig.~\ref{Figure6}(a) with the corresponding torsion profiles for trace and retrace scans presented in Fig.~\ref{Figure6}(b). The clear inversion between the trace and retrace signals suggests variations in frictional force rather than topography~\cite{jelken2022hidden}. To resolve topographic and frictional components, further analysis was carried out by evaluating the sum and difference of these signals. The difference signal ($\Delta/2$), shown as the bottom panels in Fig.~\ref{Figure6}(b), separates the frictional response by eliminating torsional components that do not reverse with scan direction whereas the sum signal ($\Sigma/2$) (Fig. S7) carries topographical information. Striped domains are clearly resolved in the LFM maps of regions R3 and R4 and exhibit higher friction force due to increased cantilever torsion. Moreover, scattered spherical particulates are also evident on the MoS$_2$ surface. These features strongly influence the lateral signal and cause local fluctuations in torsion profiles, thereby making it difficult to extract the thickness-dependent frictional variations. Further probing of the nanomechanical response was performed using FMM measurements by applying an additional sinusoidal voltage to the z-piezo at a frequency far from the tip resonant frequency~\cite{jelken2022hidden}. The topography, FMM phase, and FMM amplitude images of region R3 are shown in Fig.~\ref{Figure7}(a-c), respectively, and of R2, R4 and R5 are included in the Supporting information as Fig. S8. Topography maps clearly reveal the presence of surface particles. Corresponding line profiles shown in Fig.~\ref{Figure7}(d) indicate reduced amplitude and increased phase on MoS$_2$ compared to SiO$_2$ which implies lower effective stiffness and enhanced energy dissipation in CVD-grown MoS$_2$. Local fluctuations in the FMM amplitude and phase profiles shown in Fig.~\ref{Figure7}(d) further correlate with the observed striped patterns and surface particulates. Hence, the LFM and FMM results confirm that the striped features correspond to mechanically distinct surface domains associated with surface particulates seen in the topography. Localized charge trapping and surface dipole formation within these mechanically distinct domains give rise to the corresponding contrast revealed by KPFM. Nevertheless, the secondary contrast identified on top of the bilayer in regions R2 and R3 is not detected in either the LFM or FMM maps, suggesting that it is predominantly of electrostatic origin. 

\section{Conclusion}
In summary, MoS$_2$ bilayers exhibiting distinct morphologies associated with AA'- and AB-stacking were grown using the NaCl-assisted CVD technique. Raman and PL spectroscopy measurements confirm the bilayer nature of as-grown samples. XPS spectra reveal the presence of Na-related surface species, as evidenced by the Na 1s peak. A clear layer-dependent evolution of the work function was probed using KPFM, showing an increase in work function with layer number and a more pronounced interlayer difference in AB-stacked structures. Upon illumination, MoS$_2$ exhibits effective n-type doping driven by an enhanced photogating effect. KPFM measurements clearly resolve local surface potential variations from surface particulates and reveal their role in carrier capture under illumination thereby establishing the interplay between stacking order, substrate induced photogating and carrier trapping by surface particles in modulating the localized optoelectronic response. Nanomechanical variations arising from these spatially distributed particles were probed by LFM and FMM measurements. The present KPFM-based approach establishes an effective route to elucidate local photoresponse mechanism in TMDs with stable stacking configurations, particularly in the presence of NaCl-assisted growth-induced residues, which is essential for controlling and optimizing optoelectronic device performance.

% Experimental section

\section*{Author contributions}
\textbf{Anagha Gopinath}: Conceptualization, Investigation, Data curation, Formal analysis, Methodology, Software, Writing - original draft. \textbf{Faiha Mujeeb}: Investigation, Writing - review and editing. \textbf{Subhabrata Dhar}: Resources, Writing - review and editing. \textbf{Jyoti Mohanty}: Conceptualization, Funding acquisition, Resources, Supervision, Writing - review and editing.

\section*{Conflicts of interest}
There are no conflicts to declare.

\section*{Data availability}
The data that support the findings of this study are available from the corresponding author upon reasonable request.

\section*{Acknowledgements}
A. G. and J. M. acknowledge the Indian Institute of Technology Hyderabad for providing research facilities. A.G. acknowledges the Ministry of Education (MoE), India, for funding support. The authors also thank the Sophisticated Analytical Instrument Facility (SAIF) and the Industrial Research and Consultancy Centre (IRCC) of IIT Bombay for providing Raman and Photoluminescence characterization facilities. 

\bibliographystyle{unsrt}
\bibliography{refs}

\end{document}